\documentclass[twocolumn,aps,prl,floatfix,reprint,superscriptaddress,showpacs]{revtex4-1}
\usepackage[latin9]{inputenc}
\usepackage{color}
\usepackage{bm}
\usepackage{amsmath}
\usepackage{amssymb}
\usepackage{graphicx}
\usepackage[unicode=true,
 bookmarks=false,
 breaklinks=false,pdfborder={0 0 1},backref=false,colorlinks=true]
 {hyperref}
\hypersetup{
 letterpaper=true,pdfstartview=FitV,linkcolor=blue,citecolor=blue,urlcolor=blue}

\makeatletter

\providecommand{\tabularnewline}{\\}

\pdfoutput=1
\usepackage{amsfonts}
\usepackage{mathrsfs}
\usepackage{color}
\usepackage{bm}
\usepackage{xspace}
\usepackage{epstopdf}
\usepackage{dcolumn}
\usepackage{longtable}

\makeatother

\begin{document}

\title{An efficient method for hybrid density functional calculation with
spin-orbit coupling}

\date{\today}

\author{Maoyuan Wang}

\affiliation{Beijing Key Laboratory of Nanophotonics and Ultrafine Optoelectronic
Systems, School of Physics, Beijing Institute of Technology, Beijing
100081, China }

\affiliation{Department of physics, McGill University, Montreal, H3A 2T8, Canada}

\author{Gui-Bin Liu}
\email{gbliu@bit.edu.cn}

\affiliation{Beijing Key Laboratory of Nanophotonics and Ultrafine Optoelectronic
Systems, School of Physics, Beijing Institute of Technology, Beijing
100081, China }

\author{Hong Guo}

\affiliation{Department of physics, McGill University, Montreal, H3A 2T8, Canada}

\author{Yugui Yao}
\email{ygyao@bit.edu.cn}

\affiliation{Beijing Key Laboratory of Nanophotonics and Ultrafine Optoelectronic
Systems, School of Physics, Beijing Institute of Technology, Beijing
100081, China }
\begin{abstract}
In first-principles calculations, hybrid functional is often used
to improve accuracy from local exchange correlation functionals. A
drawback is that evaluating the hybrid functional needs significantly
more computing effort. When spin-orbit coupling (SOC) is taken into
account, the non-collinear spin structure increases computing effort
by at least eight times. As a result, hybrid functional calculations
with SOC are intractable in most cases. In this paper, we present
an approximate solution to this problem by developing an efficient
method based on a mixed linear combination of atomic orbital (LCAO)
scheme. We demonstrate the power of this method using several examples
and we show that the results compare very well with those of direct
hybrid functional calculations with SOC, yet the method only requires
a computing effort similar to that without SOC. The presented technique
provides a good balance between computing efficiency and accuracy,
and it can be extended to magnetic materials.
\end{abstract}

\pacs{63.20.dk, 73.22.-f, 71.70.Ej}
\maketitle

\section{Introduction}

Density functional theory (DFT) is a powerful method for predicting
properties of materials such as crystal structures, electronic bands,
phonon dispersions and other physical quantities. Practically, using
appropriate exchange correlation (XC) functional is very important
for accuracy, especially for predicting band gaps of materials. It
is well known that the local density approximation (LDA) and general
gradient approximation (GGA) XC functionals tend to severely underestimate
band gaps~\cite{Kohn-gap}. Consequently, hybrid functionals (HF)
such as PBE0~\cite{PBE01,PBE02,PBE03,PBE04}, HSE03 and HSE06~\cite{HSE06,HSE2,HSE3,HSE4}
were proposed and they often predict very good band gap values comparable
to experiments. A drawback of HF is that it needs very significant
computing resources, generally several orders of magnitude more compared
to that of LDA or GGA.

In recent years, materials with strong spin-orbit coupling (SOC) have
attracted great attention, including topological insulators~\cite{Hasan10,Qi11}
Bi$_{2}$Se$_{3}$~\cite{Bi2Se3}, silicene~\cite{Liu1,Liu2}, germanene~\cite{Liu1,Liu2},
stanene~\cite{Liu1,Liu2,Xu2013}, BiH~\cite{Song2014,Liu2014},
ZrTe$_{5}$~\cite{ZrTe-Weng}, Bi$_{4}$Br$_{4}$~\cite{Zhou2014},
ZrSiO~\cite{WHM-family}, photoelectric materials PbI$_{2}$~\cite{PbI2}
and BiOCl~\cite{BiOCl}, two-dimensional group-VI$_{B}$ transition
metal dichalcogenides MoS$_{2}$, MoSe$_{2}$, WS$_{2}$, and WSe$_{2}$
~\cite{LiuRev2015}, III$_{A}$-V$_{A}$ direct band-gap semiconductors
with heavy elements GaSb and InSb~\cite{k35}, etc. DFT calculations
including SOC involve non-collinear spin which requires at least eight
times more computing time as compared to that without SOC, due to
the $O(N^{3})$ scaling for solving the Kohn-Sham DFT equations (KS-DFT).
Since many of these SOC materials are semiconductors, HF calculations
are desired to more accurately predict their band gaps and electronic
structures. Unfortunately, HF+SOC calculations are numerically intractable
thus rarely used - unless the unit cell is extremely small, due to
the huge computational demand. It is the purpose of this paper to
report a practical solution to this problem.

In particular, we propose an efficient approximate technique for HF+SOC
calculations based on a mixed linear combination of atomic orbital
(LCAO) scheme. The mixed LCAO Hamiltonian is constructed by two parts:
an SOC-free part whose parameters are obtained from HF calculations
without SOC, and an SOC part whose parameters are obtained from GGA+SOC
calculations (DFT at the GGA level with SOC). Applying this approach
to several non-magnetic materials, the results are demonstrated to
be very close to those of direct HF+SOC calculation and much more
accurate than the GGA+SOC calculation. Importantly, the required computing
time of the mixed LCAO technique is comparable to that of HF calculation
without SOC.

In the rest of the work, the DFT calculations are performed using
the projector augmented wave method implemented in VASP~\cite{Kresse1996}.
The Perdew-Burke-Ernzerhof (PBE) parametrization of GGA functional~\cite{Perdew1996,KJ99}
and Heyd-Scuseria-Ernzerh hybrid functional (HSE06)~\cite{HSE06,HSE2,HSE3,HSE4}
are used in the DFT calculations, and the VASP2WANNIER90 interface~\cite{wannier2008,wannier2012,vasp2wannier90}
is used to obtain the LCAO parameters from the DFT results. Since
numerical calculations are for the purpose of demonstrating the mixed
LCAO technique, structure optimization is omitted.

\section{The method}

WANNIER90~\cite{wannier2008,wannier2012} is used to construct LCAO
or Wannier-bases Hamiltonian from DFT calculations, and the resulting
LCAO Hamiltonian can reproduce the original energy dispersion very
well. We start by constructing an LCAO Hamiltonian to treat HF+SOC
using DFT calculations.

For a given system, the required computing effort is most demanding
for HF+SOC, followed by HF without SOC and next followed by GGA+SOC.
Clearly and as explained in the Introduction, if HF+SOC were computationally
affordable in general, the work of this paper would not be necessary.
That is not the case. In the following we utilize HF without SOC and
GGA+SOC to construct a mixed LCAO Hamiltonian $H^{{\rm MIX}}$ which
we show to be a very good approximation to $H^{{\rm HF+SOC}}$. In
particular, $H^{{\rm MIX}}$ has two terms, $H_{0}^{{\rm HF}}$ which
is obtained from HF without SOC, and $H_{{\rm so}}^{{\rm GGA}}$ which
is obtained from GGA+SOC,
\begin{equation}
H^{{\rm MIX}}=H_{0}^{{\rm HF}}+H_{{\rm so}}^{\text{GGA}}.\label{Hmix0}
\end{equation}
Clearly, constructing $H^{{\rm MIX}}$ only consumes a time that is
comparable to HF without SOC, thus much more efficient than that of
a full HF+SOC calculation. The fact that $H^{{\rm MIX}}$ compare
very well with direct HF+SOC calculations (see below), suggests that
the mixed LCAO scheme provides a viable approximation for the complicated
HF+SOC analysis.

On the technical side, while $H_{0}^{{\rm HF}}$ can be constructed
directly from DFT calculation of HF without SOC, $H_{{\rm so}}^{\text{GGA}}$
is obtained from DFT of GGA+SOC involving a procedure for separating
out the SOC contributions. The latter procedure and an associated
technical detail are discussed in the following two subsections.

\subsection{Separating out the SOC contribution}

The Hamiltonian $H$ of SOC systems can be divided into a non-SOC
term $H_{0}$ plus the SOC term $H_{{\rm so}}$:
\begin{equation}
{H}=H_{0}+H_{{\rm so}}.\label{H}
\end{equation}
In LCAO representation, $H_{0}$ involves on-site energy and hopping
integral between different atomic orbitals, and $H_{{\rm so}}$ comes
from SOC effects.

In the spin-up and spin-down bases $|\!\uparrow\rangle$ and $|\!\downarrow\rangle$,
the non-SOC term $H_{0}$ can be written as a diagonal $2\times2$
matrix:
\begin{equation}
H_{0}=\left({\begin{array}{cc}
{H_{0}^{\uparrow}} & 0\\
0 & {H_{0}^{\downarrow}}
\end{array}}\right).\label{H0}
\end{equation}
For simplicity, we consider non-magnetic systems in the rest of this
work, but extension to magnetic system can be readily made without
fundamental difficulty. For non-magnetic materials, $H_{0}^{\uparrow}=H_{0}^{\downarrow}$.

For the SOC term $H_{{\rm so}}$, its original operator form is:
\begin{equation}
H_{{\rm so}}=\frac{\hbar}{{4m_{0}^{2}c^{2}}}\left(\nabla V\times\bm{p}\,\right)\cdot\bm{s}\equiv\xi\bm{\mathcal{L}}\cdot\bm{s},\label{Hso1}
\end{equation}
where $\hbar$ is the reduced Planck constant, $m_{0}$ is the bare
mass of electron, $c$ is the velocity of light, $V(\bm{r})$ is the
potential energy, $\bm{p}$ the momentum, and $\bm{s}$ the vector
of Pauli matrices representing the spin degrees of freedom. For clarity
we define a constant $\xi\equiv\hbar/(4m_{0}^{2}c^{2})$ and a vector
operator $\bm{\mathcal{L}}\equiv\nabla V\times\bm{p}$. $H_{{\rm so}}$
can then be rewritten in the following matrix form:
\begin{align}
H_{{\rm so}} & =\xi(\mathcal{L}_{x}s_{x}+\mathcal{L}_{y}s_{y}+\mathcal{L}_{z}s_{z})\nonumber \\
 & =\xi\begin{pmatrix}\mathcal{L}_{z} & \mathcal{L}_{x}-i\mathcal{L}_{y}\\
\mathcal{L}_{x}+i\mathcal{L}_{y} & -\mathcal{L}_{z}
\end{pmatrix}\equiv\left({\begin{array}{cc}
{H_{{\rm so}}^{\uparrow\uparrow}} & {H_{{\rm so}}^{\uparrow\downarrow}}\\
{H_{{\rm so}}^{\downarrow\uparrow}} & {H_{{\rm so}}^{\downarrow\downarrow}}
\end{array}}\right),\label{Hso3}
\end{align}
in which $H_{{\rm so}}^{\downarrow\downarrow}=-H_{{\rm so}}^{\uparrow\uparrow}$
and $H_{{\rm so}}^{\downarrow\uparrow}=H_{{\rm so}}^{\uparrow\downarrow\dag}$.

According to Eq.~\ref{H} to Eq.~\ref{Hso3}, the total Hamiltonian
for a non-magnetic system with SOC is:
\begin{equation}
H\equiv\begin{pmatrix}H_{11} & H_{12}\\
H_{21} & H_{22}
\end{pmatrix}=\begin{pmatrix}H_{0}^{\uparrow} & 0\\
0 & H_{0}^{\uparrow}
\end{pmatrix}+\begin{pmatrix}H_{{\rm so}}^{\uparrow\uparrow} & H_{{\rm so}}^{\uparrow\downarrow}\\
H_{{\rm so}}^{\uparrow\downarrow\dag} & -H_{{\rm so}}^{\uparrow\uparrow}
\end{pmatrix}.\label{Htot}
\end{equation}
Then, from Eq.~\ref{Htot}, we can separate the total Hamiltonian
$H$ to obtain $H_{0}$ and $H_{{\rm so}}$ as the following:
\begin{equation}
H_{0}=\begin{pmatrix}\begin{array}{cc}
\left({H_{11}+H_{22}}\right)/{2} & {0}\\
{0} & \left({H_{11}+H_{22}}\right)/{2}
\end{array}\end{pmatrix},\label{H02}
\end{equation}
\begin{equation}
H_{{\rm so}}=\begin{pmatrix}\begin{array}{cc}
\left({H_{11}-H_{22}}\right)/{2} & {H_{12}}\\
{H_{21}} & -\left({H_{11}-H_{22}}\right)/{2}
\end{array}\end{pmatrix}.\label{Hso4}
\end{equation}
Hence, after obtaining the LCAO Hamiltonian $H^{{\rm GGA+SOC}}$ from
the corresponding DFT calculation, its SOC part $H_{{\rm so}}^{{\rm GGA}}$
can be separated out using Eq.\ref{Hso4}.

\subsection{Mixing the Hamiltonian}

With the obtained non-SOC part $H_{0}^{{\rm HF}}$ and SOC part $H_{{\rm so}}^{\text{GGA}}$,
the mixed LCAO Hamiltonian $H^{{\rm MIX}}$ that approximates HF+SOC
is determined by Eq.\ref{Hmix0}. Hereinafter we use the HSE functional
for HF, and PBE functional for GGA. Then Eq.\ref{Hmix0} becomes
\begin{equation}
{H^{{\rm MIX}}}=H_{0}^{{\rm HSE}}+H_{{\rm so}}^{{\rm PBE}}.\label{Hmix}
\end{equation}
The ``mixing\char`\"{} procedure appears to be a simple addition.
However it should be noted that only when $H_{0}^{{\rm HSE}}$ and
$H_{{\rm so}}^{{\rm PBE}}$ are constructed under the same bases can
they be added directly. We achieve this by constructing $H_{0}^{{\rm HSE}}$
and $H_{{\rm so}}^{{\rm PBE}}$ in the same bases $\left|\tilde{\varphi}_{m\bm{k}}\right\rangle $,
and details are presented in the appendix \ref{sec:LCAO}. This way,
we finally constructed the mixed Hamiltonian $H^{{\rm MIX}}$ to treat
HSE+SOC.

\section{Results, analysis and discussion}

\begin{figure*}
\includegraphics[width=2\columnwidth]{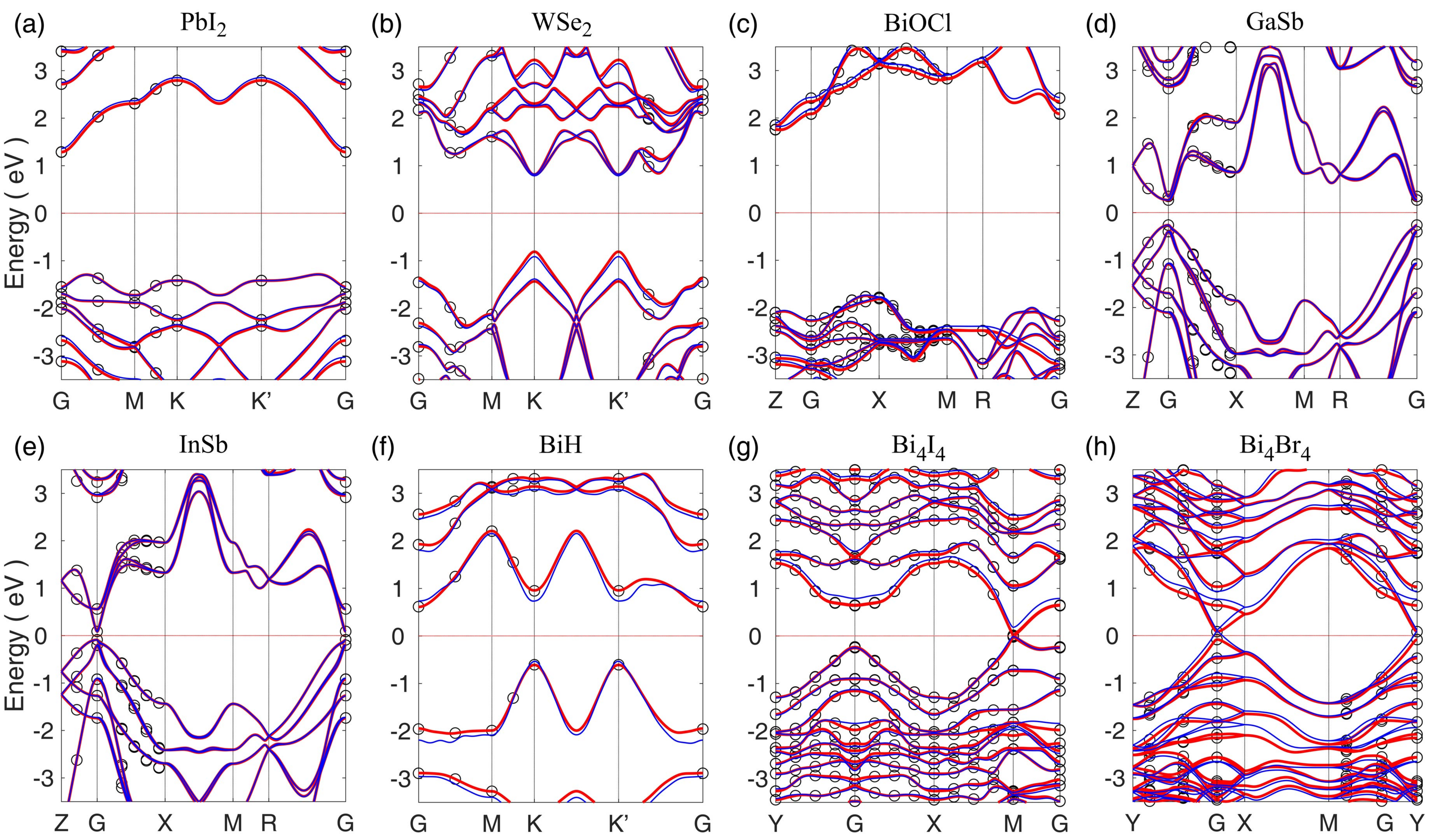} \caption{\textbf{Comparison of band structures for the example materials.}
In each band structure, black circles are the results from direct
HSE+SOC DFT calculation; red thick lines are the reproduced HSE+SOC
bands by LCAO fitting; and blue thin lines are results of our method,
i.e. $H^{{\rm MIX}}$ in Eq. \ref{Hmix}. \label{fig:band}}

\label{fig1}
\end{figure*}

Having constructed $H^{{\rm MIX}}$ to efficiently treat HSE+SOC,
in this section we demonstrate its accuracy using several material
systems. Predicting band gap is important, which is one of the reasons
to use HSE in the first place\cite{HSE06,HSE2,HSE3,HSE4}. We calculated
band gaps for eight semiconductor materials having heavy elements
thus large SOC, including two-dimensional (2D) mono-layers of PbI$_{2}$,
WSe$_{2}$, BiH, Bi$_{4}$I$_{4}$ and Bi$_{4}$Br$_{4}$; 3D crystals
BiOCl, GaSb, and InSb~\cite{k35}. The SOC effect is important for
these materials, especially for their band gaps.

For the eight materials, we performed (very time-consuming) direct
HSE+SOC calculations and using the results, we constructed an LCAO
Hamiltonian $H^{{\rm HSE+SOC}}$: this would not be possible without
the full direct HSE+SOC calculation. Then, we constructed $H^{{\rm MIX}}$
following the procedure in the last section which does not require
full HSE+SOC calculation. The three sets of results are compared:
direct numerical data from full HSE+SOC calculations and from $H^{{\rm HSE+SOC}}$,
as well as from $H^{{\rm MIX}}$.

\begin{table*}
\caption{\textbf{Comparison of band gaps }(in unit of eV) from different Hamiltonians
$H_{{\rm +SOC}}^{{\rm HSE}}$, $H^{{\rm MIX}}$, $H_{{\rm +SOC}}^{{\rm PBE}}$,
and $\tilde{H}^{{\rm MIX}}$, in which $H_{{\rm +SOC}}^{{\rm HSE(PBE)}}$
is an alternative denotion of $H^{{\rm HSE(PBE)+SOC}}$ due to the
space limit in table and $\tilde{H}^{{\rm MIX}}$ is defined as $\tilde{H}^{{\rm MIX}}=\tilde{H}_{0}^{{\rm HSE}}+H_{{\rm so}}^{{\rm PBE}}$
(for $\tilde{H}_{0}^{{\rm HSE}}$ see eq. \ref{eq:HSESOC} for details).
The band gap of a Hamiltonian $h$ is denoted as $g(h)$. $\Delta_{1}=g(H^{{\rm MIX}})-g(H_{{\rm +SOC}}^{{\rm HSE}}$),
$\Delta_{2}=g(\tilde{H}^{{\rm MIX}})-g(H_{{\rm +SOC}}^{{\rm HSE}}$),
and absolute relative deviations are also shown in percentages. For
BiH, the gap we list here is the band gap opened by SOC at the Dirac
point $K(K')$ \cite{Song2014,Liu2014}. For GaSb and InSb, PBE+SOC
calculations give wrong metallic results with no gaps. \label{tab1}}

\begin{tabular*}{1\textwidth}{@{\extracolsep{\fill}}@{\extracolsep{\fill}}c|ccccc|ccc}
\hline
\noalign{\vskip\doublerulesep}  & $g(H_{{\rm +SOC}}^{{\rm HSE}})$  & $g(H^{{\rm MIX}})$  & $g(H_{{\rm +SOC}}^{{\rm PBE}})$  & $\;\Delta_{1}$  & $|\Delta_{1}|/g(H_{{\rm +SOC}}^{{\rm HSE}})$  & $g(\tilde{H}^{{\rm MIX}})$  & $\;\Delta_{2}$  & $|\Delta_{2}|/g(H_{{\rm +SOC}}^{{\rm HSE}})$\tabularnewline
\hline
\noalign{\vskip\doublerulesep} PbI$_{2}$  & 2.569  & 2.648  & 1.862  & $\phantom{+}$0.079  & 3.08\%  & 2.626  & $\phantom{+}$0.057  & 2.22\%\tabularnewline
\noalign{\vskip\doublerulesep} WSe$_{2}$  & 1.619  & 1.702  & 1.247  & $\phantom{+}$0.083  & 5.13\%  & 1.703  & $\phantom{+}$0.084  & 5.19\%\tabularnewline
\noalign{\vskip\doublerulesep} BiOCl  & 3.496  & 3.556  & 2.511  & $\phantom{+}$0.060  & 1.72\%  & 3.502  & $\phantom{+}$0.006  & 0.17\%\tabularnewline
\noalign{\vskip\doublerulesep} GaSb  & 0.526  & 0.527  & Metal  & $\phantom{+}$0.001  & 0.19\%  & 0.531  & $\phantom{+}$0.005  & 0.95\% \tabularnewline
\noalign{\vskip\doublerulesep} InSb  & 0.175  & 0.174  & Metal  & $-$0.001  & 0.57\%  & 0.182  & $\phantom{+}$0.007  & 4.00\% \tabularnewline
\noalign{\vskip\doublerulesep} BiH  & 1.567  & 1.267  & 1.252  & $-$0.300  & 19.10\%  & 1.260  & $-$0.307  & 19.60\% \tabularnewline
\noalign{\vskip\doublerulesep} Bi$_{4}$I$_{4}$  & 0.020  & 0.206  & 0.170  & $\phantom{+}$0.186  & 930.0\%  & 0.202  & $\phantom{+}$0.182  & 910.0\%\tabularnewline
\noalign{\vskip\doublerulesep} Bi$_{4}$Br$_{4}$  & 0.166  & 0.021  & 0.357  & $-$0.145  & 87.35\%  & 0.047  & $-$0.119  & 71.69\%\tabularnewline
\hline
\end{tabular*}
\end{table*}


First, band structures of the eight compounds calculated by our method
via $H^{{\rm MIX}}$ of Eq.\ref{Hmix} is plotted in Fig.\ref{fig:band},
together with those from the direct HSE+SOC calculation and its fitting
$H^{{\rm HSE+SOC}}$. The full HSE+SOC data are presented in black
circles and the bands from $H^{{\rm HSE+SOC}}$ are in thick red lines:
these are used as benchmarks to compare to our results by $H^{{\rm MIX}}$
which are presented in thin blue lines. We can see that the bands
calculated by our method via $H^{{\rm MIX}}$ of Eq.\ref{Hmix} (thin
blue lines) are qualitatively consistent to the benchmark results
for all cases. In particular, band dispersions by $H^{{\rm MIX}}$
and the benchmark $H^{{\rm HSE+SOC}}$ agree well for PbI$_{2}$,
WSe$_{2}$, GaSb, and InSb; and the agreement is somewhat reduced
for BiOCl, BiH, Bi$_{4}$I$_{4}$, and Bi$_{4}$Br$_{4}$. In the
latter cases, although the heavy element Bi gives rise to some quantitative
difference, there is no qualitative discrepancy for the full range
of the Brillouin zone.

Second, quantitatively we compare the band gaps of $H^{{\rm MIX}}$
and the benchmark $H^{{\rm HSE+SOC}}$ in Table \ref{tab1}. The band
gaps obtained by $H^{{\rm MIX}}$ are close to those of the benchmark
$H^{{\rm HSE+SOC}}$ for five of the eight compounds PbI$_{2}$, WSe$_{2}$,
BiOCl, GaSb, and InSb {[}see the column labeled by $|\Delta_{1}|/g(H_{{\rm +SOC}}^{{\rm HSE}})$
column in Table \ref{tab1}{]}. But for three compounds BiH, Bi$_{4}$I$_{4}$
and Bi$_{4}$Br$_{4}$ - especially Bi$_{4}$I$_{4}$ which has a
very small band gap, the discrepancy is large. As a supplement, we
also show the calculated band gaps by PBE+SOC in Table \ref{tab1}:
they are not only quantitatively quite different from the benchmark
results, two of them are even qualitatively wrong (GaSb, InSb).

As for the three compounds with band gaps of $H^{{\rm MIX}}$ showing
large discrepancy to the benchmark, BiH, Bi$_{4}$I$_{4}$ and Bi$_{4}$Br$_{4}$,
they are all relevant to topological insulators which exhibit band
inversion near the Fermi level when SOC is considered and have normal
bands with SOC not considered. Concretely, 2D monolayers of BiH and
Bi$_{4}$Br$_{4}$ are topological insulators\cite{Zhou2014}. As
for Bi$_{4}$I$_{4}$ monolayer, although it is not a topological
insulator, it lies near the transition point between normal and topological
insulator which makes it also sensitive to SOC. The impact of topological
properties on the accuracy of our method needs to be analyzed.

Recall that we used $H^{{\rm MIX}}=H_{0}^{{\rm HSE}}+H_{{\rm so}}^{{\rm PBE}}$
to approximate $H^{{\rm HSE+SOC}}$ which can be written as
\begin{equation}
H^{{\rm HSE+SOC}}=\tilde{H}_{0}^{{\rm HSE}}+H_{{\rm so}}^{{\rm HSE}}\label{eq:HSESOC}
\end{equation}
where $\tilde{H}_{0}^{{\rm HSE}}$ and $H_{{\rm so}}^{{\rm HSE}}$
were obtained directly from $H^{{\rm HSE+SOC}}$ using Eq.\ref{H02}
and Eq.\ref{Hso4}, respectively. Therefore, any discrepancy between
$H^{{\rm MIX}}$ and $H^{{\rm HSE+SOC}}$ has two sources: (i) the
discrepancy between the non-SOC part $H_{0}^{{\rm HSE}}$ (obtained
from HSE without SOC) and $\tilde{H}_{0}^{{\rm HSE}}$; (ii) the discrepancy
between the SOC part $H_{{\rm so}}^{{\rm PBE}}$ and $H_{{\rm so}}^{{\rm HSE}}$.
We analyze these terms in the next two subsections.

\subsection{HSE without SOC}


In this subsection we analyze the source of discrepancy in the non-SOC
part $H_{0}^{{\rm HSE}}$ (obtained from HSE without SOC) and $\tilde{H}_{0}^{{\rm HSE}}$.
Since DFT is based on ground state electronic density, any discrepancy
should be due to the difference between densities calculated by the
two approaches \cite{HSE-WF}. To illustrate this, we proceed by making
use of a model Hamiltonian as follows:
\begin{align}
H= & \sum\limits _{c}{\varepsilon_{0c}a_{c}^{\dag}a_{c}}+\sum\limits _{v}{\varepsilon_{0v}a_{v}^{\dag}a_{v}}\nonumber \\
 & +\sum\limits _{c,c^{\prime}}{\xi_{c,c^{\prime}}^{{\rm SO}}a_{c}^{\dag}a_{c^{\prime}}}+\sum\limits _{v,v^{\prime}}{\xi_{v,v^{\prime}}^{{\rm SO}}a_{v}^{\dag}a_{v^{\prime}}}\label{HTI}\\
 & +\sum\limits _{c,v}{(\xi_{c,v}^{{\rm SO}}}a_{c}^{\dag}a_{v}+{\rm h.c.})\nonumber
\end{align}
where the first and second terms are the non-SOC part $H_{0}$, the
third to fifth terms \textemdash{} denoted as $H_{{\rm c,c}}^{{\rm SO}}$,
$H_{{\rm v,v}}^{{\rm SO}}$, and $H_{{\rm c,v}}^{{\rm SO}}$ respectively,
constitute the SOC part $H_{{\rm so}}$. In Eq.\ref{HTI}, $a_{c(v)}$
and $a_{c(v)}^{\dag}$ are respectively annihilation and creation
operators for the $c$ conduction ($v$ valence) band\cite{Add1}
eigen-state $\psi_{0c}$ ($\psi_{0v}$) of $H_{0}$ whose eigenenergy
is $\varepsilon_{0c}$ ($\varepsilon_{0v}$).

\begin{figure}
\includegraphics[width=1\columnwidth]{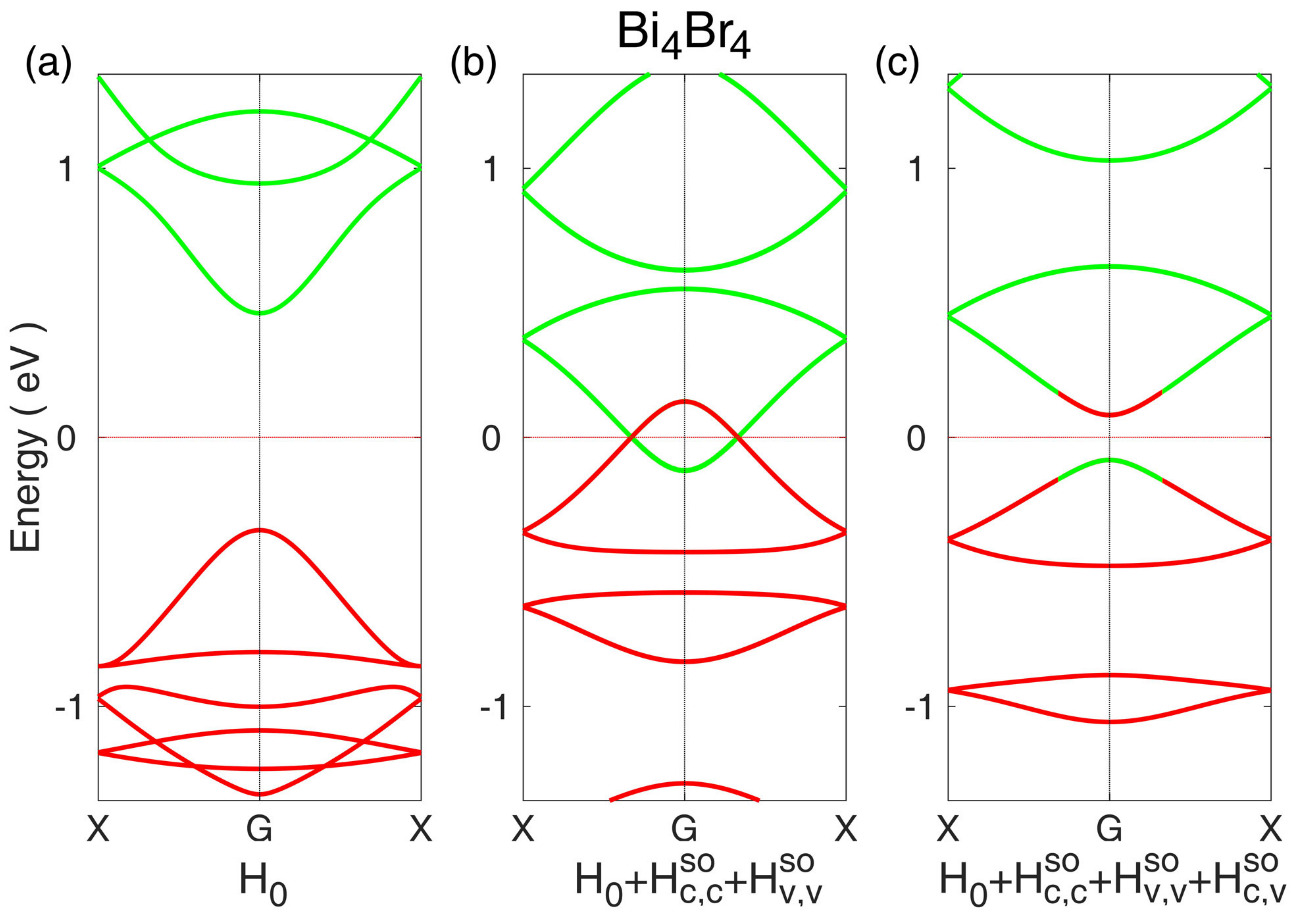} \caption{\textbf{Type-I and type-II SOC effects in an HSE+SOC DFT calculation.}
Here Bi$_{4}$Br$_{4}$ is taken as example. (a)-(c) are the band
structures of different Hamiltonians. (a) Bands of the Hamiltonian
without SOC ($H_{0}$), where the red (green) lines represent the
valence (conduction) band composition of $H_{0}$. (b) Bands perturbed
by the strong type-I SOC term, which results in band splittings that
make valence (conduction) bands cross Fermi energy. (c) Bands in (b)
perturbed further by the type-II SOC term, in which case the gap opens
again. Here it is obvious that charge density is changed from (a)
to (c), because the valence bands in (c) mix some parts of the previous
conduction (green) bands in (a).}

\label{fig2}
\end{figure}

The Hamiltonian in Eq.\ref{HTI} can be analyzed by perturbation theory.
We take the non-SOC $H_{0}$ as the unperturbed Hamiltonian and the
SOC term $H_{{\rm so}}$ as the perturbation. Note that $H_{{\rm so}}=H_{{\rm c,c}}^{{\rm SO}}+H_{{\rm v,v}}^{{\rm SO}}+H_{{\rm c,v}}^{{\rm SO}}$
can be divided into two types according to their SOC effects: $H_{{\rm c,c}}^{{\rm SO}}$
and $H_{{\rm v,v}}^{{\rm SO}}$ are the ``type-I\char`\"{} terms,
$H_{{\rm c,v}}^{{\rm SO}}$ is the ``type-II\char`\"{} term. The
type-I term $H_{{\rm c,c}}^{{\rm SO}}+H_{{\rm v,v}}^{{\rm SO}}$ couples
conduction bands with other conduction bands as well as valence bands
with other valence bands. We distinguish two situations. (i) If the
unperturbed band gap is large relative to the type-I SOC effect, the
type-I term only splits bands and no band crossing occurs. This is
the ``weak type-I SOC'' which does not make any significant difference
between the unperturbed and perturbed charge densities (see Appendix
\ref{sec:appB} for details). (ii) If the unperturbed band gap is
small relative to the type-I SOC effect, some split bands near the
band gap will cross the Fermi level so that the charge density is
altered. This is the ``strong type-I SOC'' {[}cf. Fig.~\ref{fig2}(b){]}.

As for the type-II SOC $H_{{\rm c,v}}^{{\rm SO}}$, it mixes valence
bands with conduction bands, for instance the first-order perturbed
valence band state $\psi_{v}$ is
\begin{equation}
\psi_{v}\approx\frac{1}{C}(\psi_{0v}+\sum\limits _{c}{\alpha_{c}\psi_{0c}}),\label{eq:psiv-pert}
\end{equation}
in which $\alpha_{c}=\xi_{c,v}^{{\rm SO}}/(\varepsilon_{0v}-\varepsilon_{0c})$
and $C$ is a normalization constant. As a result, the type-II SOC
term alters charge density via the sum of all occupied states, i.e.
all valence band states in the presence of band gap {[}cf. Fig.~\ref{fig2}(c){]}.

\begin{table}
\caption{Major orbital compositions of the highest valence band and the lowest
conduction band for each example material. The orbital compositions
listed for WSe$_{2}$ are only for band edges at the $K$($K'$) point.
The p orbital means the collective of p$_{x}$, p$_{y}$, and p$_{z}$
orbitals. Bi$_{{\rm in}}$ and Bi$_{{\rm ex}}$ mean different Bi
atoms located at interior and exterior positions respectively.}

\label{Table.VC} %
\begin{tabular*}{1\columnwidth}{@{\extracolsep{\fill}}@{\extracolsep{\fill}}c|cc}
\hline
 & valence band  & conduction band\tabularnewline
\hline
PbI$_{2}$  & I-p  & Pb-p\tabularnewline
WSe$_{2}$ \cite{Liu2013}  & W-d$_{xy}$\&d$_{x^{2}-y^{2}}$  & W-d$_{z^{2}}$\tabularnewline
BiOCl  & O-p\&Cl-p  & Bi-p\tabularnewline
GaSb  & Sb-p  & Ga-p\tabularnewline
InSb  & Sb-p  & In-p\tabularnewline
BiH \cite{Liu2014}  & Bi-p$_{x,y}$  & Bi-p$_{x,y}$\tabularnewline
Bi$_{4}$I$_{4}$ \cite{Zhou2014}  & Bi$_{{\rm in}}$-p$_{x}$\&Bi$_{{\rm ex}}$-p$_{x}$  & Bi$_{{\rm in}}$-p$_{x}$\&Bi${\rm _{ex}}$-p$_{x}$\tabularnewline
Bi$_{4}$Br$_{4}$\cite{Zhou2014}  & Bi$_{{\rm in}}$-p$_{x}$\&Bi$_{{\rm ex}}$-p$_{x}$  & Bi$_{{\rm in}}$-p$_{x}$\&Bi$_{{\rm ex}}$-p$_{x}$\tabularnewline
\hline
\end{tabular*}
\end{table}

Let us first understand why for the five normal band insulators PbI$_{2}$,
BiOCl, WSe$_{2}$, GaSb and InSb, $H^{{\rm MIX}}$ agrees with the
benchmark $H^{{\rm HSE+SOC}}$ very well. For these compounds, the
properties of their lowest conduction band differ significantly from
those of highest valence band, especially in orbital compositions
and positions (cf. Table.\ref{Table.VC}). This makes the type-II
SOC hopping $\xi_{c,v}^{{\rm SO}}$ composed mostly of the SOC interactions
between different atoms, and hence $\xi_{c,v}^{{\rm SO}}$ are very
small. WSe$_{2}$ is an exception because its band-edge orbitals locate
at the same atom W. However the W-d$_{xy}$\&d$_{x^{2}-y^{2}}$ orbitals
have opposite mirror symmetry compared to W-d$_{z^{2}}$ orbital,
which still makes $\xi_{c,v}^{{\rm SO}}$ small. Furthermore, the
gaps of the normal band insulators are large compared to $\xi_{c,v}^{{\rm SO}}$,
i.e. $\left|\varepsilon_{0v}-\varepsilon_{0c}\right|\text{\ensuremath{\gg}}\left|\xi_{c,v}^{{\rm SO}}\right|$.
Consequently the perturbed valence band eigenfunction $\psi_{v}$
can at most slightly mix with the conduction band $\psi_{0c}$ since
$\alpha_{c}\ll1$ (see Eq. \ref{eq:psiv-pert}), which means type-II
SOC has almost no effect. This is shown in table \ref{Table.Hcv}
which illustrates band gaps of $H-H_{{\rm c,v}}^{{\rm SO}}$ and $H$
differ by less than 2.3\%. In addition, their band gaps are also large
enough compared to the type-I SOC hopping $\xi_{v,v'}^{{\rm SO}}$
and $\xi_{c,c'}^{{\rm ,SO}}$, hence this is the weak type-I SOC case
which makes no change of charge density if type-II SOC is not considered.
We conclude that for normal band insulators, both type-I and type-II
SOC effects contribute very little change to the charge density from
$\rho_{0}=\sum_{v}\psi_{v0}^{*}\psi_{v0}$ to $\rho=\sum_{v}\psi_{v}^{*}\psi_{v}$.
This is why that in DFT calculations of these compounds, $H_{0}$
of HSE without SOC ($H_{0}^{{\rm HSE}}$) and HSE+SOC ($\tilde{H}_{0}^{{\rm HSE}}$)
are very close to each other since charge densities are very close.

\begin{table}
\caption{\textbf{Gap comparison of two Hamiltonians differing by $H_{{\rm c,v}}^{{\rm SO}}$
(in unit of eV).} $H-H_{{\rm c,v}}^{{\rm SO}}$ means the total Hamiltonian
with full SOC $H$ excluding $H_{{\rm c,v}}^{{\rm SO}}$, which equals
$H_{0}+H_{{\rm c,c}}^{{\rm SO}}+H_{{\rm v,v}}^{{\rm SO}}$. $\Delta_{{\rm cv}}=g(H)-g(H-H_{{\rm c,v}}^{{\rm SO}})$.
Gaps listed here are calculated in HSE+SOC case to exemplify the slight
effects of the type-II SOC \textbf{$H_{{\rm c,v}}^{{\rm SO}}$} in
the normal band insulators. }

\label{Table.Hcv} %
\begin{tabular*}{1\columnwidth}{@{\extracolsep{\fill}}@{\extracolsep{\fill}}c|cccc}
\hline
 & $g(H-H_{{\rm c,v}}^{{\rm SO}})$  & $g(H)$  & $\Delta_{{\rm cv}}$  & $|\Delta_{{\rm cv}}|/g(H)$ \tabularnewline
\hline
PbI$_{2}$  & 2.554  & 2.569  & 0.015  & 0.58\%\tabularnewline
WSe$_{2}$  & 1.604  & 1.619  & 0.015  & 0.93\% \tabularnewline
BiOCl  & 3.442  & 3.496  & 0.054  & 1.54\% \tabularnewline
GaSb  & 0.521  & 0.526  & 0.005  & 0.95\% \tabularnewline
InSb  & 0.171  & 0.175  & 0.004  & 2.29\% \tabularnewline
\hline
\end{tabular*}
\end{table}

\begin{table*}
\caption{\textbf{Gap comparison of Hamiltonians without SOC (in unit of eV):
$H_{0}^{{\rm HSE}}$ vs $\tilde{H}_{0}^{{\rm HSE}}$ as well as $H_{0}^{{\rm PBE}}$
vs $\tilde{H}_{0}^{{\rm PBE}}$.} Similar to $\tilde{H}_{0}^{{\rm HSE}},$
we define $\tilde{H}_{0}^{{\rm PBE}}=H^{{\rm PBE+SOC}}-H_{{\rm so}}^{{\rm PBE}}$
which means the non-SOC part separated out from the PBE+SOC Hamiltonian.
$\Delta_{0}^{{\rm HSE}}=g(H_{0}^{{\rm HSE}})-g(\tilde{H}_{0}^{{\rm HSE}})$
and $\Delta_{0}^{{\rm PBE}}=g(H_{0}^{{\rm PBE}})-g(\tilde{H}_{0}^{{\rm PBE}}$).
Note that BiH without SOC shows Dirac-cone type of bands with no gaps
for comparison, and InSb is calculated to be metallic in PBE without
SOC and does not have gaps either. }

\label{Table.H0} %
\begin{tabular*}{1\textwidth}{@{\extracolsep{\fill}}@{\extracolsep{\fill}}c|cccc|cccc}
\hline
\noalign{\vskip\doublerulesep}  & $g(\tilde{H}_{0}^{{\rm HSE}})$  & $g(H_{0}^{{\rm HSE}})$  & $\Delta_{0}^{{\rm HSE}}$  & $|\Delta_{0}^{{\rm HSE}}|/g(\tilde{H}_{0}^{{\rm HSE}})$  & $g(\tilde{H}_{0}^{{\rm PBE}})$  & $g(H_{0}^{{\rm PBE}})$  & $\Delta_{0}^{{\rm PBE}}$  & $|\Delta_{0}^{{\rm PBE}}|/g(\tilde{H}_{0}^{{\rm PBE}})$ \tabularnewline
\hline
PbI$_{2}$  & 3.279  & 3.286  & $\text{\ensuremath{\phantom{+}}}$0.007  & 0.21\%  & 2.529  & 2.517  & $-$0.012  & 0.47\%\tabularnewline
WSe$_{2}$  & 1.996  & 1.995  & $-$0.001  & 0.05\%  & 1.554  & 1.547  & $-$0.007  & 0.45\% \tabularnewline
BiOCl  & 3.722  & 3.773  & $\text{\ensuremath{\phantom{+}}}$0.051  & 1.37\%  & 2.753  & 2.782  & $\text{\ensuremath{\phantom{+}}}$0.029  & 1.05\% \tabularnewline
GaSb  & 0.773  & 0.769  & $-$0.004  & 0.52\%  & 0.123  & 0.114  & $-$0.009  & 7.32\% \tabularnewline
InSb  & 0.429  & 0.421  & $-$0.008  & 1.86\%  & Metal  & Metal  & -  & - \tabularnewline
BiH  & Dirac  & Dirac  & -  & -  & Dirac  & Dirac  & -  & - \tabularnewline
Bi$_{4}$I$_{4}$  & 1.050  & 1.085  & $\text{\ensuremath{\phantom{+}}}$0.035  & 3.33\%  & 0.657  & 0.610  & $-$0.047  & 7.15\% \tabularnewline
Bi$_{4}$Br$_{4}$  & 0.806  & 0.771  & $-$0.035  & 4.34\%  & 0.408  & 0.283  & $-$0.125  & 30.64\% \tabularnewline
\hline
\end{tabular*}
\end{table*}

Next, we analyze the three compounds where $H^{{\rm MIX}}$ has significant
discrepancy to the benchmark $H^{{\rm HSE+SOC}}$. As discussed above,
these compounds have inverted bands (BiH, Bi$_{4}$Br$_{4}$) or near-inverted
bands (Bi$_{4}$I$_{4}$), the orbital compositions of their conduction
bands are similar to those of valence bands around the Fermi energy
(cf. Table.\ref{Table.VC}). Hence $\left|\xi_{c,v}^{{\rm SO}}\right|$
is not small in general and can be close to or even larger than $|\varepsilon_{0v}-\varepsilon_{0c}|$.
According to Eq.\ref{eq:psiv-pert}, this makes $\alpha_{c}$ not
small and the type-II SOC significantly perturb $\psi_{v}$. In addition,
the large SOC strength of Bi can lead to a strong type-I SOC that
make conduction and valence bands cross the Fermi energy {[}see Fig.~\ref{fig2}(b){]}.
Due to the strong type-I and type-II SOC effects, the charge density
changes significantly from $\rho_{0}$ to $\rho$, which makes $H_{0}^{{\rm HSE}}$
and $\tilde{H}_{0}^{{\rm HSE}}$ differ from each other.

Quantitatively, we compare the band gaps of $H_{0}^{{\rm HSE}}$ and
$\tilde{H}_{0}^{{\rm HSE}}$ in Table.~\ref{Table.H0}, from which
we observe that the gaps of $H_{0}^{{\rm HSE}}$ are very close to
those of $\tilde{H}_{0}^{{\rm HSE}}$ for PbI$_{2}$, BiOCl, WSe$_{2}$,
GaSb and InSb, but not so close for Bi$_{4}$I$_{4}$ and Bi$_{4}$Br$_{4}$.
As a comparison, we also show the gaps of $H_{0}^{{\rm PBE}}$ and
$\tilde{H}_{0}^{{\rm PBE}}=H_{0}^{{\rm PBE+SOC}}-H_{{\rm so}}^{{\rm PBE}}$
in Table \ref{Table.H0}. It should be pointed out that because the
PBE gaps without SOC are smaller than HSE gaps, i.e. $|\varepsilon_{v0}-\varepsilon_{c0}|$
are smaller in PBE, $\psi_{v}$ will mix more $\psi_{0c}$ if $\xi_{c,v}^{{\rm SO}}$
and $\rho_{0}$ are assumed to be the same in HSE and PBE. Hence the
gap differences of PBE are worse than those of HSE for GaSb, Bi$_{4}$I$_{4}$
and Bi$_{4}$Br$_{4}$, especially for Bi$_{4}$Br$_{4}$, as shown
in table \ref{Table.H0}.

\subsection{PBE with SOC}


Having understood the discrepancy in the non-SOC part $H_{0}^{{\rm HSE}}$
(obtained from HSE without SOC) and $\tilde{H}_{0}^{{\rm HSE}}$,
in this subsection we analyze the discrepancy between the SOC part
$H_{{\rm so}}^{{\rm PBE}}$ and $H_{{\rm so}}^{{\rm HSE}}$. For different
XC functionals such as PBE and HSE, there are two major reasons that
make $H_{{\rm so}}$ different in PBE and HSE.

\begin{figure*}
\includegraphics[width=2\columnwidth]{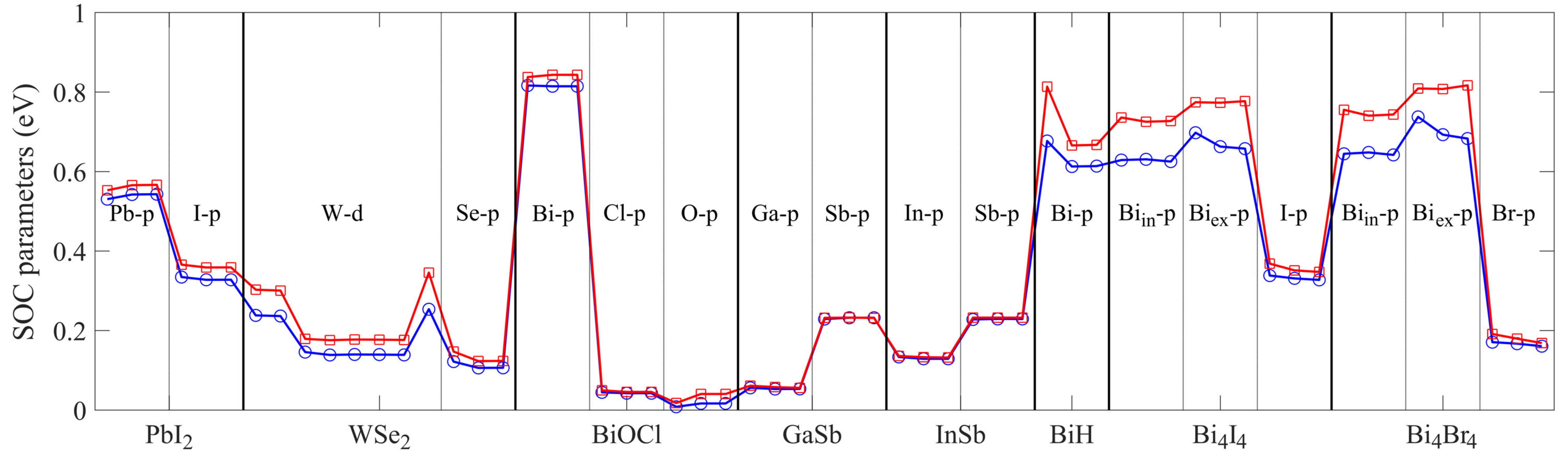} \caption{\textbf{On site SOC parameters in Hamiltonian.} The blue circles (red
squares) are from PBE+SOC (HSE+SOC), and lines are provided to guide
the eye. For p-orbitals, three circles (squares) mean $\left|\left\langle p_{x}^{\uparrow}\left|\widehat{H}_{{\rm so}}\right|p_{y}^{\uparrow}\right\rangle \right|$,
$\left|\left\langle p_{y}^{\uparrow}\left|\widehat{H}_{{\rm so}}\right|p_{z}^{\downarrow}\right\rangle \right|$
and $\left|\left\langle p_{z}^{\uparrow}\left|\widehat{H}_{{\rm so}}\right|p_{x}^{\downarrow}\right\rangle \right|$
respectively. For d-orbitals of $W$ in the figure, eight circles
(squares) mean $\left|\left\langle d_{z^{2}}^{\uparrow}\left|\widehat{H}_{{\rm so}}\right|d_{xz}^{\downarrow}\right\rangle \right|$,
$\left|\left\langle d_{z^{2}}^{\uparrow}\left|\widehat{H}_{{\rm so}}\right|d_{yz}^{\downarrow}\right\rangle \right|$,
$\left|\left\langle d_{xz}^{\uparrow}\left|\widehat{H}_{{\rm so}}\right|d_{yz}^{\uparrow}\right\rangle \right|$,
$\left|\left\langle d_{xz}^{\uparrow}\left|\widehat{H}_{{\rm so}}\right|d_{x^{2}-y^{2}}^{\downarrow}\right\rangle \right|$,
$\left|\left\langle d_{xz}^{\uparrow}\left|\widehat{H}_{{\rm so}}\right|d_{xy}^{\downarrow}\right\rangle \right|$,
$\left|\left\langle d_{yz}^{\uparrow}\left|\widehat{H}_{{\rm so}}\right|d_{x^{2}-y^{2}}^{\downarrow}\right\rangle \right|$,
$\left|\left\langle d_{yz}^{\uparrow}\left|\widehat{H}_{{\rm so}}\right|d_{xy}^{\downarrow}\right\rangle \right|$,
and $\left|\left\langle d_{x^{2}-y^{2}}^{\uparrow}\left|\widehat{H}_{{\rm so}}\right|d_{xy}^{\uparrow}\right\rangle \right|$
respectively. Bi$_{{\rm in}}$ and Bi$_{{\rm ex}}$ mean Bi atoms
in different positions~\cite{Zhou2014}. It should be pointed out
that the orbitals here are the orthonormalized orbitals (see eq. \ref{LCAO9}
in Appendix \ref{sec:LCAO}).}

\label{fig3}
\end{figure*}

First, according to Eq.~\ref{Hso1}, different potential energy $V$
makes different SOC parameters. For HSE, its exchange potential $V_{{\rm x}}^{{\rm HSE}}$
is composed of a short-ranged part $V_{{\rm x}}^{{\rm SR}}$ and a
long-ranged part $V_{{\rm x}}^{{\rm LR}}$, where $V_{{\rm x}}^{{\rm SR}}$
is produced by mixing the non-local Fock potential $V_{{\rm x}}^{{\rm F}}$
(i.e. the exact exchange potential) and the PBE exchange potential
$V_{{\rm x}}^{{\rm PBE}}$ in short range, while $V_{{\rm x}}^{{\rm LR}}$
is solely the PBE exchange potential $V_{{\rm x}}^{{\rm PBE}}$ in
long range. Since the unscreened Fock potential $V_{{\rm x}}^{{\rm F}}$
is generally larger than the PBE exchange potential, the resulting
$V_{{\rm x}}^{{\rm HSE}}$ with $V_{{\rm x}}^{{\rm PBE}}$ replaced
partially by the unscreened $V_{{\rm x}}^{{\rm F}}$, should also
be larger than $V_{{\rm x}}^{{\rm PBE}}$ in general. This is demonstrated
by the SOC hopping parameters of HSE and PBE in Fig.~\ref{fig3},
in which the HSE ones are consistently larger than the PBE ones. Taking
BiH as an example: it has a hexagonal structure like graphane and
has Dirac cones at $K$ and $K'$ points in the Brillouin zone without
SOC. With SOC, a topological band gap opens at $K\ (K')$ point and
this gap depends only on the strength of SOC. Hence, the larger gap
of HSE+SOC than that of our method shown in Fig\ref{fig1}(f) means
that the SOC strength of HSE+SOC is larger than that of our method,
i.e. $H_{{\rm so}}^{{\rm HSE}}$ is larger than $H_{{\rm so}}^{{\rm PBE}}$.
For other topological insulators similar to BiH, i.e. the ones which
have Dirac points or node lines before including SOC, such as ZrTe$_{5}$\cite{ZrTe-Weng}
and ZrSiO families\cite{WHM-family} which have band inversion before
considering SOC just as BiH does, they share similar source of discrepancy
to BiH in our method\cite{Add2}. Second, if $\xi_{c,v}^{{\rm SO}}$
and $\rho_{0}$ of PBE and HSE are assumed to be the same, using the
perturbation theory, for near-inverse band insulator (Bi$_{4}$I$_{4}$)
or inverse band topological insulator (Bi$_{4}$Br$_{4}$), different
gaps between PBE and HSE make $\psi_{v}$ mix at different ratios
with $\psi_{0c}$, resulting in different charge densities which gives
further differences of $V$ and $H_{{\rm so}}$ between PBE and HSE.

To quantitatively understand the difference between $H_{{\rm so}}^{{\rm HSE}}$
and $H_{{\rm so}}^{{\rm PBE}}$, we compare the gaps of two Hamiltonians:
one is $H^{{\rm HSE+SOC}}$, and the other is $H^{{\rm HSE+SOC}}$
with its SOC part $H_{{\rm so}}^{{\rm HSE}}$ replaced by $H_{{\rm so}}^{{\rm PBE}}$,
i.e. $\tilde{H}^{{\rm MIX}}$ in Table \ref{tab1}. We can see from
Table \ref{tab1} (especially the last column) that the gap differences
induced by the difference between $H_{{\rm so}}^{{\rm HSE}}$ and
$H_{{\rm so}}^{{\rm PBE}}$ are relatively small for band insulators
PbI$_{2}$, BiOCl, WSe$_{2}$, GaSb and InSb, but large for BiH, Bi$_{4}$I$_{4}$
and Bi$_{4}$Br$_{4}$.

According to these analyses, for normal band insulators, the differences
between $H_{0}^{{\rm HSE}}$ and $\tilde{H}_{0}^{{\rm HSE}}$ are
as small as the differences between $H_{{\rm so}}^{{\rm HSE}}$ and
$H_{{\rm so}}^{{\rm PBE}}$. This is why our method works so wy our method works so well
for the five compounds PbI$_{2}$, WSe$_{2}$, BiOCl, GaSb and InSb
(see Table \ref{tab1}). But for the three near-inverse and inverse
band insulators BiH, Bi$_{4}$I$_{4}$ and Bi$_{4}$Br$_{4}$, the
differences between $H_{{\rm so}}^{{\rm HSE}}$ and $H_{{\rm so}}^{{\rm PBE}}$
are much larger {[}cf. the $|\Delta_{0}^{{\rm HSE}}|/g(\tilde{H}_{0}^{{\rm HSE}})$
column in table \ref{Table.H0} and the $|\Delta_{2}|/g(H_{{\rm +SOC}}^{{\rm HSE}})$
column in table \ref{tab1}{]}. Furthermore, by comparing $\Delta_{1}$
and $\Delta_{2}$ in Table \ref{tab1}, we conclude that the error
of our method is dominated by the difference between $H_{{\rm so}}^{{\rm HSE}}$
and $H_{{\rm so}}^{{\rm PBE}}$, and this error are large for the
near-inverse and inverse band insulators. Note, however, although
the \emph{relative} deviation of our $H^{{\rm MIX}}$ with respect
to $H^{{\rm HSE+SOC}}$ is large for Bi$_{4}$I$_{4}$ and Bi$_{4}$Br$_{4}$,
their \emph{absolute} deviations are actually not large, as illustrated
by the not-so-large differences of the SOC hopping parameters between
PBE and HSE shown in Fig.\ref{fig3}.

We therefore conclude that our method is a very good approximation
for normal band insulators and, for near-inverse or inverse band insulators
which have very small band gaps, our method is still reasonable in
that it can provide qualitatively correct results.

\subsection{Discussions}

The purpose of this work is to develop a reasonably accurate, qualitatively
correct and computationally efficient method to perform HSE+SOC calculations
within DFT. We have so far demonstrated the accuracy of our method
and understood the source of discrepancy when dealing with near-inverse
and inverse band insulators.

Concerning computational efficiency: our method for HSE+SOC calculation
takes essentially the same time as an HSE calculation without SOC.
Taking WSe$_{2}$ for example, in our calculations, one PBE+SOC electronic
step takes 1.3 minute using 32 CPU cores, one HSE (without SOC) electronic
step takes 5.6 minute using 128 CPU cores, and one full HSE+SOC electronic
step takes 50.5 minute using 128 CPU cores. Therefore our technique
is nearly an order of magnitude faster than the full HSE+SOC calculation.
We checked that for all the cases we investigated, our method is faster
by several times to more than an order of magnitude that the full
direct HSE+SOC approach.

So far we analyzed non-magnetic compounds, but the method can be easily
extended to magnetic materials for which $H_{0}^{\uparrow}$ and $H_{0}^{\downarrow}$
are not equal anymore but can be obtained in an additional spin-colinear
DFT calculation. Namely, our method can be extended to magnetic materials
by performing a spin-colinear PBE calculation and a PBE+SOC calculation
to extract $H_{{\rm so}}^{{\rm PBE}}$; then performing a spin-colinear
HSE calculation to obtain $H_{0}^{{\rm HSE}}$. The results are added
together to obtain $H^{{\rm MIX}}$ for the magnetic material.

The method developed in this work is not only suitable for LCAO, but
also useful for accelerating HSE+SOC DFT calculations. Usually, two
initializations are used to save computing time during HSE+SOC calculations:
(a) using charge density and wave function from a PBE+SOC calculation
as initialization, (b) using charge density and wave function from
a HSE calculation without SOC as initialization. However, both will
actually not accelerate calculation significantly. This is because
for (a), $H_{0}$ of PBE+SOC differ quite a lot from that of HSE+SOC;
and for (b), it lacks $H_{{\rm so}}$. Then, naturally, our method
provides a better starting charge density and wave functions for full
HSE+SOC DFT calculation because $H^{{\rm MIX}}$ is closer to $H^{{\rm HSE+SOC}}$
than $H^{{\rm PBE+SOC}}$ or $H_{0}^{{\rm HSE}}$.

\section{Summary}

In summary, we have developed an efficient mixed LCAO technique to
perform HSE+SOC DFT calculations. The LCAO Hamiltonian is obtained
by mixing a non-SOC part and an SOC part. The non-SOC part is constructed
by SOC-free HSE, and the SOC part by PBE+SOC. As a result, the mixed
LCAO technique requires a computing time comparable to that of a non-SOC
HSE calculation, thus saving about one order magnitude in computing
time compared to a full direct HSE+SOC calculation.

Applying the method to eight non-magnetic compounds demonstrates that
the mixed LCAO Hamiltonian can well approximate that of the full HSE+SOC.
In particular, the method works very well for normal band insulators,
and it is also reasonable to give qualitatively correct results for
near-inverse and inverse band insulators having very small band gaps.
We find that the errors in our method came from the difference between
$H_{{\rm so}}^{{\rm HSE}}$ and $H_{{\rm so}}^{{\rm PBE}}$ more than
the difference between $\tilde{H}_{0}^{{\rm HSE}}$ and $H_{0}^{{\rm HSE}}$
in most cases. Our method can be easily extended to other hybrid functionals
and magnetic materials, and it can also be used to provide good initial
conditions for full direct HSE+SOC calculation.

\section{Acknowledgments}

The authors thank Qing Shi of Mcgill University for the data of GaSb
and InSb. The work is supported by the MOST Project of China (Grant
No. 2014CB920903), the NSF of China (Grant Nos. 11734003, 11574029),
the National Key R\&D Program of China (Grant No. 2016YFA0300600)
(MW and YY); by the National Key R\&D Program of China with Grant
No. 2017YFB0701600 and NSFC with Grant No. 11304014(GBL); and Natural
Science and Engineering Research Council (NSERC) of Canada (HG). HG
thanks Compute Canada for computational facilities where part of this
work was carried out.

\appendix

\section{\label{sec:LCAO}LCAO representation and interpolation}

Using VASP2WANNIER90~\cite{wannier2008,wannier2012,vasp2wannier90},
we can project the local orbitals ${g_{n}(\bm{r})}$ onto the Bloch
manifold $\psi_{m\bm{k}}$ determined by VASP at wave vector $\bm{k}$
to obtain
\begin{equation}
\left|{\varphi_{n\bm{k}}}\right\rangle \equiv\sum\limits _{m=1}^{N_{b}}{\left|{\psi_{m\bm{k}}}\right\rangle \left\langle {\psi_{m\bm{k}}}\right|\left.{g_{n}}\right\rangle },\label{LCAO1}
\end{equation}
\begin{equation}
A_{mn}^{\bm{k}}\equiv\left\langle {\psi_{m\bm{k}}}\right|\left.{g_{n}}\right\rangle ,\label{LCAO2}
\end{equation}
where $g_{n}$ is localized trial orbitals serving as the initial
guess of the Wannier functions, and $N_{b}$ is the number of bands
considered or the dimension of the Bloch manifold at $\bm{k}$. Thus
obtained $\left|{\varphi_{n\bm{k}}}\right\rangle $ are only determined
by the local orbitals ${g_{n}}$, for $\left|{\varphi_{n\bm{k}}}\right\rangle $
can be Fourier transformed to $g_{n}^{\bm{R}}(\bm{r})\equiv g_{n}(\bm{r}-\bm{R})$
in which $\bm{R}$ is lattice vector. To show this, first apply the
Bloch theorem $\varphi_{n\bm{k}}(\bm{r}-\bm{R})=e^{-i\bm{k}\cdot\bm{R}}\varphi_{n\bm{k}}(\bm{r})$
to eq. \ref{LCAO1} to get
\begin{equation}
\left|{\varphi_{n\bm{k}}}\right\rangle =e^{i\bm{k}\cdot\bm{R}}\sum\limits _{m=1}^{N_{b}}{\left|{\psi_{m\bm{k}}}\right\rangle \left\langle {\psi_{m\bm{k}}}\right|\left.{g_{n}^{\bm{R}}}\right\rangle },
\end{equation}
and then do Fourier transformation to get
\begin{multline}
\quad\frac{1}{N}\sum_{\bm{k}}{e^{-i\bm{k}\cdot\bm{R}}\left|{\varphi_{n\bm{k}}}\right\rangle }\\
=\frac{1}{N}\sum\limits _{m\bm{k}}{\left|{\psi_{m\bm{k}}}\right\rangle \left\langle {\psi_{m\bm{k}}}\right|\left.{g_{n}^{\bm{R}}}\right\rangle }=\left|g_{n}^{\bm{R}}\right\rangle \quad\label{gnRn}
\end{multline}
where $N$ is the number of unit cells (also the number of $\bm{k}$
points). In the above equation, we have used the completeness relation
\begin{equation}
\frac{1}{N}\sum_{m\bm{k}}\left|\psi_{m\bm{k}}\right\rangle \left\langle \psi_{m\bm{k}}\right|=1\label{sumpsi}
\end{equation}
due to the normalization convention $\left\langle \psi_{m\bm{k}}\right|\left.\psi_{n\bm{k}'}\right\rangle =N\delta_{mn}\delta_{\bm{kk}'}$
\cite{wannier2012}. The inverse transformation of eq. \ref{gnRn}
shows that $\left|{\varphi_{nk}}\right\rangle $ is just the Bloch
sum of $g_{n}$:
\begin{equation}
\left|{\varphi_{n\bm{k}}}\right\rangle =\sum_{\bm{R}}e^{i\bm{k}\cdot\bm{R}}g_{n}^{\bm{R}}.\label{Blochsum}
\end{equation}

Under the bases of $\left|\varphi_{n\bm{k}}\right\rangle $, the Hamiltonian
matrix is $\mathcal{H}^{\bm{k}}=A^{\bm{k}\dag}\mathcal{E}^{\bm{k}}A^{\bm{k}}$
with matrix elements
\begin{align}
\begin{array}{l}
\mathcal{H}_{ij}^{\bm{k}}\end{array} & =\left\langle {\varphi_{i\bm{k}}}\right|\hat{H}\left|{\varphi_{j\bm{k}}}\right\rangle \nonumber \\
 & =\frac{1}{N^{2}}\sum\limits _{m,n}{\left\langle {\varphi_{i\bm{k}}}\right|\left.{\psi_{m\bm{k}}}\right\rangle \left\langle {\psi_{m\bm{k}}}\right|\hat{H}\left|{\psi_{n\bm{k}}}\right\rangle \left\langle {\psi_{n\bm{k}}}\right|\left.{\varphi_{j\bm{k}}}\right\rangle }\nonumber \\
 & =\sum_{m,n}A_{im}^{\bm{k}\dag}\mathcal{E}_{mn}^{\bm{k}}A_{nj}^{\bm{k}}\label{Hnor1}
\end{align}
in which $\mathcal{E}_{mn}^{\bm{k}}=\delta_{mn}\varepsilon_{n\bm{k}}$
and $\varepsilon_{n\bm{k}}$ is the eigenenergy of the Bloch state
$|\psi_{n\bm{k}}\rangle$. In the above equation, eq. \ref{sumpsi}
and the relation $\left\langle {\psi_{n\bm{k}}}\right|\left.{\varphi_{j\bm{k}}}\right\rangle =NA_{nj}^{\bm{k}}$
(see eq. \ref{LCAO1} and \ref{LCAO2}) are used. Because $g_{n}^{\bm{R}}$
and hence $\left|\varphi_{n\bm{k}}\right\rangle $ are not orthonormalized,
the eigen equation of $\mathcal{H}^{\bm{k}}$ is

\begin{equation}
\mathcal{H}^{\bm{k}}\phi=\varepsilon S^{\bm{k}}\phi\label{Hnor2}
\end{equation}
where the overlap matrix $S^{\bm{k}}$ is defined by
\begin{equation}
S_{ij}^{\bm{k}}=\frac{1}{N}\left\langle {\varphi_{i\bm{k}}}\right|\left.{\varphi_{j\bm{k}}}\right\rangle =\left({A^{\bm{k}\dag}}{A^{\bm{k}}}\right)_{ij}.\label{Sij}
\end{equation}
To orthonormalize the bases, we construct $\tilde{\varphi}_{n\bm{k}}$
with properties $\left\langle \tilde{\varphi}_{m\bm{k}}\right|\left.\tilde{\varphi}_{n\bm{k}}\right\rangle =N\delta_{mn}$
as follows
\begin{equation}
\left|{\tilde{\varphi}_{n\bm{k}}}\right\rangle =\sum\limits _{m=1}^{N_{b}}\left|{\varphi_{m\bm{k}}}\right\rangle T_{mn}^{\bm{k}}\label{LCAO4}
\end{equation}
in which $T^{\bm{k}}$ is a Hermitian matrix with the property $(T^{\bm{k}})^{2}=(S^{\bm{k}})^{-1}$
and hence can be denoted as $T^{\bm{k}}=(S^{\bm{k}})^{-\frac{1}{2}}$
in form. The existence of $T^{\bm{k}}$ is guaranteed by the Hermiticity
of $S^{\bm{k}}$. As a result, the Hamiltonian matrix under the orthonormalized
bases $|\tilde{\varphi}_{m\bm{k}}\rangle$ is
\begin{equation}
H(\bm{k})=T^{\bm{k}\dag}\mathcal{H}^{\bm{k}}T^{\bm{k}}=(A^{\bm{k}}T^{\bm{k}})^{\dag}\mathcal{E}^{\bm{k}}A^{\bm{k}}T^{\bm{k}},\label{LCAO5}
\end{equation}
which can be constructed from the data $\varepsilon_{m\bm{k}}$ and
$A_{mn}^{\bm{k}}$ generated by VASP and VASP2WANNIER90 respectively.

Because $\varphi_{m\bm{k}}$ are determined only by $g_{n}$ (eq.
\ref{Blochsum}), $S^{\bm{k}}$ is determined only by $\varphi_{m\bm{k}}$
(eq. \ref{Sij}), $T^{\bm{k}}$ is determined only by $S^{\bm{k}}$,
and $\tilde{\varphi}_{m\bm{k}}$ are determined only by $\varphi_{m\bm{k}}$
and $T^{\bm{k}}$ (eq. \ref{LCAO4}), it can be concluded that the
final bases $\tilde{\varphi}_{m\bm{k}}$ are only determined by the
initial local orbitals $g_{n}$ and irrelevant to the Bloch states
$\psi_{m\bm{k}}$. This is crucial for us to add directly $H_{0}^{{\rm HSE}}$
and $H_{{\rm so}}^{{\rm PBE}}$ constructed from independent DFT calculations,
because $H_{0}^{{\rm HSE}}$ and $H_{{\rm so}}^{{\rm PBE}}$ have
the same bases $\tilde{\varphi}_{m\bm{k}}$ as long as the same initial
local orbitals $g_{n}$ are used.

The obtained $H(\bm{k}$) from eq. \ref{LCAO5} is defined at only
a finite number ($N$) of $\bm{k}$ points. If the Hamiltonian at
an arbitrary wave vector $\bm{q}$ different from the $N$ $\bm{k}$
points is required, interpolation has to be done. The interpolation
can be achieved through two steps. First, do a Fourier transformation
for $H$($\bm{k}$) to get the Hamiltonian element in real space
\begin{equation}
H_{nm}^{\bm{R}}=\left\langle {\textbf{0}n}\right|\widehat{H}\left|{\bm{R}m}\right\rangle =\frac{1}{N}\sum\limits _{\bm{k}}{e^{-i\bm{k}\cdot\bm{R}}H_{nm}(\bm{k})}.\label{LCAO7}
\end{equation}
in which $|\bm{R}m\rangle$ is the Fourier transformation of $\left|{\tilde{\varphi}_{m\bm{k}}}\right\rangle $
\begin{equation}
\left|{\bm{R}m}\right\rangle =\frac{1}{N}\sum\limits _{\bm{k}}{e^{-i\bm{k}\cdot\bm{R}}\left|{\tilde{\varphi}_{m\bm{k}}}\right\rangle }\label{LCAO9}
\end{equation}
and is an orthonormalized local orbital with property $\left\langle \bm{R}n\right|\left.\bm{R}'m\right\rangle =\delta_{\bm{RR}'}\delta nm$.
Then, construct the Hamiltonian at arbitrary wave vector $\bm{q}$
using the $\bm{k}$-indepnedent quantities $H_{nm}^{\bm{R}}$ as follows
:
\begin{equation}
H_{nm}(\bm{q})=\sum\limits _{\bm{R}}{e^{i\bm{q}\cdot\bm{R}}H_{nm}^{\bm{R}}}.\label{LCAO8}
\end{equation}
We call this procedure LCAO interpolation.

\section{\label{sec:appB}The weak type-I SOC effect}

To interpret the effect of the type-I SOC, we consider the Hamiltonian
eq. \ref{HTI} in the absence of type-II SOC hoppings $\xi_{c,v}^{{\rm SO}}$
\begin{multline}
H=\sum\limits _{c}{\varepsilon_{0c}a_{c}^{\dag}a_{c}}+\sum\limits _{v}{\varepsilon_{0v}a_{v}^{\dag}a_{v}}\\
+\sum\limits _{c,c^{\prime}}{\xi_{c,c^{\prime}}^{{\rm SO}}a_{c}^{\dag}a_{c^{\prime}}}+\sum\limits _{v,v^{\prime}}{\xi_{v,v^{\prime}}^{{\rm SO}}a_{v}^{\dag}a_{v^{\prime}}},\label{Hnocv}
\end{multline}
in which $v,v'=1,\cdots,N_{{\rm v}}$ represent valence bands, $c,c'=N_{{\rm v}}+1,\cdots,N_{{\rm b}}$
represent conduction bands, $N_{{\rm v}}$ and $N_{{\rm c}}$ are
the number of valence and conduction bands respectively, and $N_{{\rm b}}=N_{{\rm v}}+N_{{\rm c}}$
is the total number of bands considered. Note that the spin index
is incorporated into the band index $v$ and $c$ here for simplicity.
Choosing the eigen states $\psi_{0c/v}$ of $H_{0}$ as bases, $H_{0}$
is a diagonal matrix $H_{0}={\rm diag}\{H_{0}^{{\rm v}},H_{0}^{{\rm c}}\}$
with
\begin{equation}
H_{0}^{{\rm v}}={\rm diag}\{\varepsilon_{01},\cdots,\varepsilon_{0N_{{\rm v}}}\},
\end{equation}
\begin{equation}
H_{0}^{{\rm c}}={\rm diag}\{\varepsilon_{0,N_{{\rm v}}+1},\cdots,\varepsilon_{0N_{{\rm b}}}\},
\end{equation}
and $H_{{\rm so}}$ is a block-diagonal matrix $H_{{\rm so}}={\rm diag}\{H_{{\rm so}}^{{\rm v}},H_{{\rm so}}^{{\rm c}}\}$
with $H_{{\rm so}}^{{\rm v}}$ and $H_{{\rm so}}^{{\rm c}}$ being
$N_{{\rm v}}\times N_{{\rm v}}$ and $N_{{\rm c}}\times N_{{\rm c}}$
matrices respectively. Hence, the charge density of $H_{0}$ is

\begin{equation}
{\rho_{0}}=\sum_{v=1}^{N_{{\rm v}}}\left|{\psi_{0v}}\right|^{2}=\sum_{v=1}^{N_{{\rm v}}}\left|{\psi_{v}}\right|^{2},\label{rho0}
\end{equation}
where $\psi_{v}$ is the eigen state of $H_{{\rm v}}=H_{0}^{{\rm v}}+H_{{\rm so}}^{{\rm v}}$
and the second equality is due to the fact that $\psi_{v}$ is related
to $\psi_{0v}$ by a unitary transformation $U$ with $\psi_{v}=\sum_{v'}\psi_{0v'}U_{v'v}$
and $U_{v'v}=\left\langle \psi_{0v'}\right|\left.\psi_{v}\right\rangle $
(For simplicity we use the usual normalization convention $\left\langle \psi_{0v}\right|\left.\psi_{0v'}\right\rangle =\left\langle \psi_{v}\right|\left.\psi_{v'}\right\rangle =\delta_{vv'}$
here). This is guaranteed by the absence of type-II SOC hoppings $\xi_{c,v}^{{\rm SO}}$.

If the type-I SOC effect is weak, ${H_{{\rm so}}}$ will not make
conduction or valence bands cross the Fermi energy like Fig.~\ref{fig2}(b).
This makes the charge density of $H=H_{0}+H_{{\rm so}}$ is just determined
by the valence bands $\psi_{v}$
\begin{equation}
\rho=\sum_{v=1}^{N_{{\rm v}}}\left|{\psi_{v}}\right|^{2}.
\end{equation}
Considering eq. \ref{rho0} we have
\begin{equation}
\rho=\rho_{0}.
\end{equation}
Therefore, the weak type-I SOC effect cannot make difference between
the charge densities $\rho$ and $\rho_{0}$.


\begin{thebibliography}{10}
\bibitem{HSE-WF} Since HSE has non-local Fock potential included,
which depends on ground state wave function, the difference between
ground state wave functions calculated by the two approaches will
also cause the discrepancy. However, ground state wave function plays
a similar role of ground state electronic density, so we do not discuss
it individually.
\bibitem{Add2} Calculations show that the band-structure discrepancies
introduced in our method for ZrTe$_{5}$ and ZrSiO are obviously smaller
than for BiH due to smaller SOC strength of Zr and Te atoms than Bi
atom.
\end{thebibliography}
\end{document}